\documentclass[conference]{IEEEtran}
\IEEEoverridecommandlockouts
\usepackage{cite}
\usepackage{amsmath,amssymb,amsfonts}
\usepackage{algorithmic}
\usepackage{graphicx}
\usepackage{textcomp}
\usepackage{xcolor}
\usepackage{tabularx}
\usepackage{booktabs}
\ifCLASSOPTIONcompsoc
    \usepackage[caption=false, font=normalsize, labelfont=sf, textfont=sf]{subfig}
\else
\usepackage[caption=false, font=footnotesize]{subfig}

\usepackage{tikz}
\usetikzlibrary{arrows}

\begin{document}

\title{Space-Time Phase Coupling in STMM-based Wireless Communications}

\author{Marouan~Mizmizi\IEEEauthorrefmark{1},  Dario~Tagliaferri\IEEEauthorrefmark{1}, Marco Di Renzo\IEEEauthorrefmark{2} and Umberto~Spagnolini\IEEEauthorrefmark{1}\\
\IEEEauthorblockA{\IEEEauthorrefmark{1}Politecnico di Milano, Dipartimento di Elettronica, Informazione e Bioingegneria, Italy \\
\IEEEauthorrefmark{2}Université Paris-Saclay, CNRS, CentraleSupélec, Laboratoire des Signaux et Systèmes, France\\
Emails: \{marouan.mizmizi,dario.tagliaferri,umberto.spagnolini\}@polimi.it, marco.di-renzo@universite-paris-saclay.fr}
\thanks{This work was partially supported by the European Union under the Italian National Recovery and Resilience Plan (NRRP) of NextGenerationEU, partnership on “Telecommunications of the Future” (PE00000001 - program “RESTART”).
}
\thanks{The work of M. Di Renzo was supported in part by the European Commission through the H2020 ARIADNE project under grant agreement number 871464 and through the H2020 RISE-6G project under grant agreement number 101017011, and by the Agence Nationale de la Recherche (ANR) through the ANR PEPR-5G project.}
}

\maketitle

\begin{abstract}

Space-time modulated metasurfaces (STMMs) are a recently proposed generalization of reconfigurable intelligent surfaces, which include a proper time-varying phase at the metasurface elements, enabling higher flexibility and control of the reflected signals. The spatial component can be designed to control the direction of reflection, while the temporal one can be adjusted to change the frequency of the reflected signal or to convey information. However, the coupling between the spatial and temporal phases at the STMM can adversely affect its performance. Therefore, this paper analyzes the system parameters that affect the space-time coupling. Furthermore, two methods for space-time decoupling are investigated. Numerical results highlight the effectiveness of the proposed decoupling methods and reveal that the space-time phase coupling increases with the bandwidth of the temporal phase, the size of the STMM, and with grazing angles of incidence onto the STMM.

\end{abstract}


\section{Introduction}

Electromagnetic (EM) metasurfaces, a.k.a. reconfigurable intelligent surfaces (RISs), are commonly regarded as one of the enablers of 6G services, being the key component of the \textit{smart EM environment} \cite{di2020smart}. RISs are made of sub-wavelength-sized elements, referred to as meta-atoms, whose scattering properties can be engineered to manipulate the impinging EM waves and control the angles of reflection and refraction \cite{9999292}. 

Recent research advances proposed RISs (and metasurfaces in general) as a candidate technology in many applications, including channel blockage mitigation \cite{Zeng9539048} \cite{Mizmizi2022_conformal}, coverage extension \cite{Kurt9359529} and, more recently, as an enhancement for sensing~\cite{9732186}. The RIS technology is based on the \textit{generalized Snell's law}, formulated in 2011 \cite{doi:10.1126/science.1210713}, which asserts that setting a spatial phase gradient across the RIS elements breaks the spatial symmetry (and thus the conservation of the linear momentum of the reflected and refracted waves). In other words, an EM wave impinging from an arbitrary angle $\theta_i$ can be reflected toward an angle $\theta_o\neq \theta_i$. The relation between $\theta_i$ and $\theta_o$ is a deterministic function of the spatial phase gradient. More recently, in 2014, the authors of \cite{shaltout2015time, IJ133_NatComm_9_4334_2018} proposed the \textit{universal Snell's law}, extending the generalized one by considering also a temporal phase gradient applied at each element of the RIS. The latter operation allows the RIS to break the Lorentz reciprocity and enables anomalous nonreciprocal reflections. In this case, an EM wave at carrier frequency $f_i$ impinging on a time-modulated RIS from angle $\theta_i$ can be reflected towards an angle $\theta_o\neq \theta_i$ and with a frequency $f_o\neq f_i$. This opens the possibility to realize new devices, including space-time modulated metasurfaces (STMMs), that not only control the direction of reflection but also change the frequency and convey information \cite{taravati2022microwave}.



The authors of \cite{GengBo_STCM2022} apply a temporally periodic phase gradient at each element of the RIS, which generates temporal and spatial harmonics, i.e., the impinging signal is reflected in different directions with different frequency shifts depending on the set periodicity. A similar approach is proposed by the authors of \cite{9133266}, where the generated harmonics encode the symbols of a quadrature amplitude modulation. 
The idea was validated by using a demonstrator capable of establishing a 20 Mbps communication link. The authors of \cite{8901437} discuss the realization of advanced modulation schemes by engineering the phase response of the single STMM element. In this direction, the work \cite{9499059} analyzes the capacity of a RIS-aided link, where information data are jointly encoded at the transmitter and at the RIS. The authors prove that the latter scheme is necessary to achieve the maximum information rate and provides significant gain over the max-signal-to-noise ratio approach, which fixes the reflection pattern. A system-level analysis is shown in \cite{9516949}, with a focus on a beamforming-plus-phase information system toward a single user. Finally, the authors of \cite{9473842} propose to apply phase-modulated STMM to enhance backscatter communications. 

All the aforementioned works are focused on the physical realization of STMM or discuss specific applications. This paper analyzes the coupling between the spatial and temporal phases that can be applied at the STMM. The space-time phase coupling has not been considered in the literature so far (see e.g., \cite{9133266}), and manifests itself when the incident signal is not perpendicular to the STMM and when the phase-modulated reflection coefficient of the STMM has a relatively large bandwidth compared to the size of the STMM (e.g., for information-bearing signals). Specifically, we consider a generic full-duplex system where a master unit (MU) sends a downlink signal of bandwidth $B_d$ to a slave unit (SU), equipped with a STMM. The latter is used to \textit{back-reflect and modulate} the received MU $\rightarrow$ SU signal with a MU $\leftarrow$ SU uplink signal of bandwidth $B_u$. The uplink signal is encoded in the STMM reflection coefficient's phase, reducing the energy consumption of the SU \cite{9749219}. In this context, the contributions of the paper are twofold: \textit{(i)} we analytically characterize the coupling between the spatial and temporal phases in STMM-based wireless systems, generalizing our previous work \cite{Mizmizi2023_STMM} by considering a hybrid STMM structure, whereby a variable number of wideband tunable components are employed; \textit{(ii)} we evaluate the impact of the spatial-temporal phase coupling on the achievable spectral efficiency of the MU $\leftarrow$ SU back-reflection link. Specifically, we show the existence of a cut-off bandwidth limit in the back-reflection MU $\leftarrow$ SU link and introduce proper \textit{decoupling} countermeasures. 

\textit{Organization}: The remainder of the present paper is organized as follows. Section \ref{sect:SysModel} describes the universal Snell's law and introduces the system and channel model. Section \ref{sec:STCDesign} outlines the principles of the considered STMM temporal modulation, highlighting the space-time phase coupling and proposing suitable decoupling methods. Section \ref{sect:FDComm} reports numerical simulations to illustrate the performance of the proposed solutions. Section \ref{sect:conclusion} concludes the paper.

\textit{Notation}: Bold upper- and lower-case letters denote matrices and column vectors. The $(i,j)$-th entry of matrix $\mathbf{A}$ is denoted by $[\mathbf{A}]_{i,j}$. Matrix transposition, conjugate transposition, and Frobenius norm are denoted as $\mathbf{A}^{T}$, $\mathbf{A}^{H}$ and $\|\mathbf{A}\|_F$, respectively. $\mathrm{diag}(\mathbf{A})$ denotes the extraction of the diagonal of $\mathbf{A}$, while $\mathrm{diag}(\mathbf{a})$ is the diagonal matrix whose elements on the main diagonal are those of vector $\mathbf{a}$. $\mathbf{I}_n$ denotes the identity matrix of size $n$.  $\mathbf{a}\sim\mathcal{CN}(\boldsymbol{\mu},\mathbf{C})$ denotes a multi-variate circularly complex Gaussian random variable with mean $\boldsymbol{\mu}$ and covariance $\mathbf{C}$. $\mathbb{E}[\cdot]$ is the expectation operator, * denotes the convolution operator in the time domain, and $\mathbb{R}$ and $\mathbb{C}$ stand for the set of real and complex numbers, respectively. $\delta(t)$ is the Dirac delta function in time and $\delta_{n}$ denotes the Kronecker delta function.


\begin{figure}[t!]
    \centering
    
    \includegraphics[width=\columnwidth]{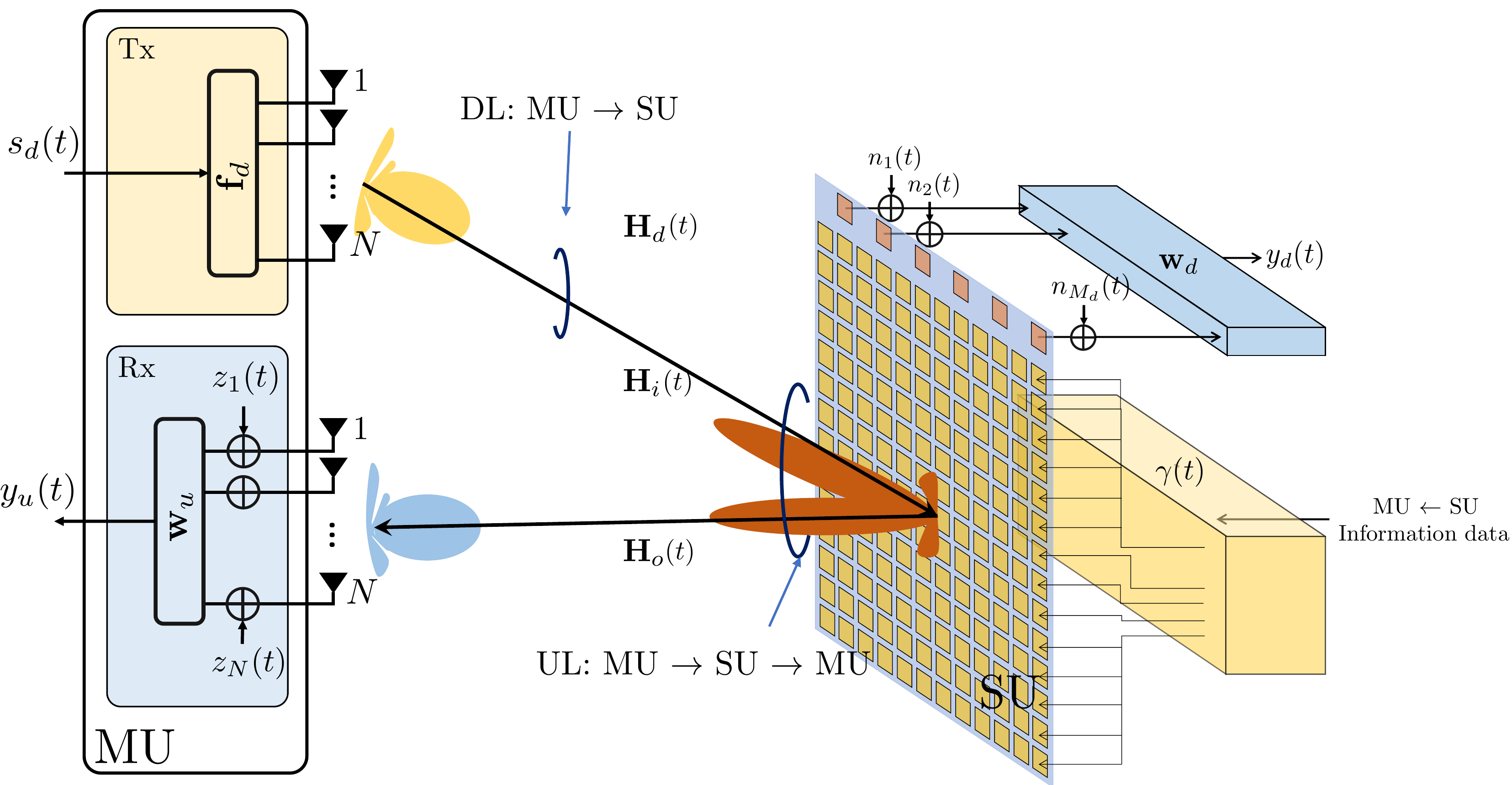}
    \caption{Full-duplex wireless communication system enabled by an STMM.}
    \label{fig:refScen}
\end{figure}

\section{System Model with STMM}\label{sect:SysModel}

\begin{figure}[b!]
\vspace{-.5cm}
    \centering
    \includegraphics[width=0.6\columnwidth]{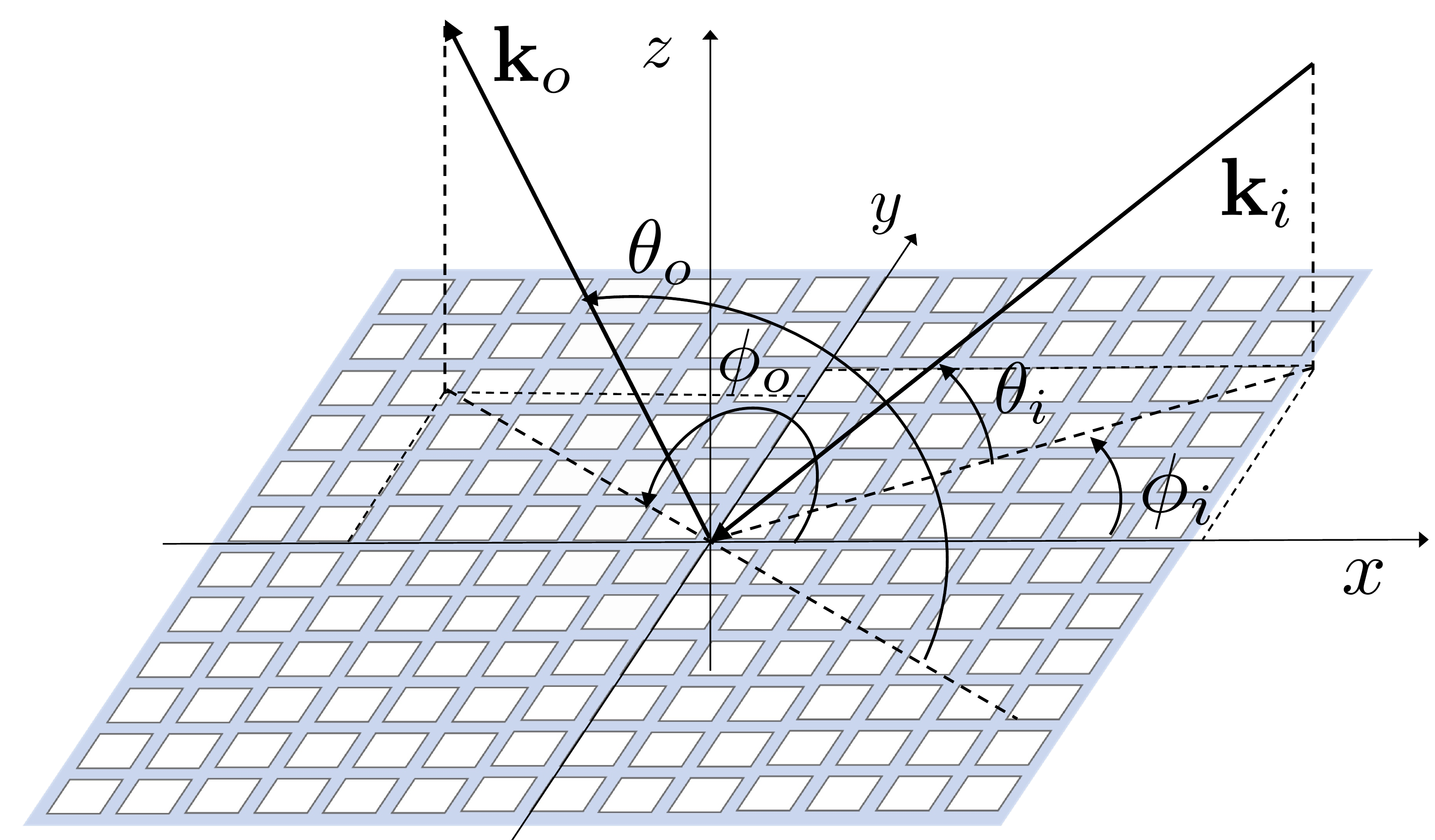}
    \caption{Reference system.}
    \label{fig:refSystem}
\end{figure}
%
 
Let us consider the wireless communication system depicted in Fig. \ref{fig:refScen}, where an MU establishes a full-duplex link with an SU. The MU is equipped with two antenna arrays of $N$ elements each, while the SU is equipped with a planar STMM of $M_u=M_{u,x}\times M_{u,y}$ elements, with $M_{u,x}$ and $M_{u,y}$ denoting the number of elements on the $x$-axis and $y$-axis, respectively, and a receiving (Rx) antenna array of $M_d$ elements. The SU Rx is meant to receive and decode the MU $\rightarrow$ SU signal, while the STMM is utilized to modulate data onto the MU $\leftarrow$ SU link via back-reflection.

\subsection{The Universal Snell's Law and the STMM}\label{subsect:STMM}

The universal Snell's law is a generalization of the generalized Snell's law when dealing with both spatial and temporal phase gradients across a metasurface. 
Let us consider the scheme depicted in Fig. \ref{fig:refSystem}, where a plane EM wave at carrier frequency $f_i$ impinges on a planar STMM from direction $(\phi_i,\theta_i)$, whose wavevector is denoted by $\mathbf{k}_i$.
The STMM, deployed on the $xy$ plane, is characterized by a space-time varying reflection coefficient
\begin{equation}\label{eq:rm}
    \Gamma(\mathbf{x}, \boldsymbol{\vartheta},t) = \alpha(\boldsymbol{\vartheta}) \; e^{j\beta(\mathbf{x},t)},
\end{equation}
where $\mathbf{x}=[x,y]^T$ is the generic position across the STMM and $\beta(\mathbf{x},t)$ denote the time- and space-varying phase of the reflection coefficient, and $\alpha(\boldsymbol{\vartheta})$ denotes the amplitude of the reflection coefficient, which depends on the incident and reflected angles $\boldsymbol{\vartheta} = (\phi_i,\theta_i, \phi_o,\theta_o)$. 
The reflected wave is characterized by the wavevector $\mathbf{k}_o$ and the carrier frequency $f_o$, which, based on the universal Snell's law, depends on the incident EM wave as follows \cite{IJ133_NatComm_9_4334_2018}:
\begin{align}
    \mathbf{k}_o - \mathbf{k}_i &= \left[\frac{\beta(\mathbf{x},t)}{\partial{x}},\frac{\beta(\mathbf{x},t)}{\partial{y}}, 0\right]^\mathrm{T}, \label{eq:spaceGradient}\\
    f_o - f_i &= \frac{1}{2\pi} \frac{\beta(\mathbf{x},t)}{\partial{t}}.\label{eq:timeGradient}
\end{align}
We note that a non-zero spatial gradient of the phase $\beta(\mathbf{x},t)$ changes the direction of reflection specified by the wavevector $\mathbf{k}_o$, while a non-zero temporal gradient of the phase $\beta(\mathbf{x},t)$ changes the carrier frequency $f_o$ of the reflected EM wave. 

\subsection{Signal Model}
The time-continuous signal from the MU can be expressed as 
\begin{equation}\label{eq:Tx_signal_w_precoding}
    \mathbf{x}(t) = \mathbf{f}_d\;s_d(t)e^{j2\pi f_i t}
\end{equation}
where $\mathbf{f}_d\in\mathbb{C}^{N\times 1}$ is the spatial precoding vector and $s_d(t)$ is the MU $\rightarrow$ SU time-continuous baseband information signal with bandwidth $B_d$ and centered at the carrier frequency $f_i$. 
%
%
The signal received at the SU can be expressed as
\begin{equation}\label{eq:dlSignal}
\begin{split}
    y_d(t) & = \frac{1}{\sqrt{\varrho_d}} \, \mathbf{w}_d^H \mathbf{H}_d(t)* \mathbf{x}(t) + \mathbf{w}_d^H \mathbf{n}(t)
\end{split}
\end{equation}
where $\varrho_d$ represents the MU $\rightarrow$ SU power path loss,  $\mathbf{w}_d\in\mathbb{C}^{N\times 1}$ is the SU spatial combiner, $\mathbf{H}_d(t)\in \mathbb{C}^{M_d \times N}$ is the MU $\rightarrow$ SU MIMO channel matrix, such that $\mathrm{E}\left[\|\mathbf{H}_d(t)\|_\mathrm{F}^2\right] = N M_d$ while  $\mathbf{n}(t)\sim\mathcal{CN}(0,\sigma^2_{n}\mathbf{I}_{M_d})$ is the additive noise, here assumed as temporally uncorrelated.

The back-reflected signal at the MU can be formulated as
\begin{equation}\label{eq:ulSignal}
\begin{split}
       y_u(t) = \frac{1}{\sqrt{\varrho_u}} \mathbf{w}_u^H \mathbf{H}_o(t) * \boldsymbol{\Gamma}(\boldsymbol{\vartheta},t)\left(\mathbf{H}_i(t) * \mathbf{x}(t) \right) + \mathbf{w}_u^H \mathbf{z}(t)
\end{split}
\end{equation}
where $\varrho_u$ is the MU $\rightarrow$ SU $\rightarrow$ MU power path loss, $\mathbf{w}_u\in\mathbb{C}^{N\times 1}$ is the combiner at the MU, $\mathbf{H}_i(t)\in\mathbb{C}^{M_u\times N}$ is the \textit{forward} MIMO channel between the MU and the STMM, $\mathbf{H}_o(t)\in\mathbb{C}^{N\times M_u}$ is the \textit{backward} MIMO channel between the STMM and the MU, $\mathbf{z}(t)\sim\mathcal{CN}(0,\sigma^2_{z} \mathbf{I}_N)$ is the additive noise, and $\boldsymbol{\Gamma}(\boldsymbol{\vartheta}, t) \in \mathbb{C}^{M_u \times M_u}$ denotes the time-varying STMM reflection coefficient matrix
\begin{equation}\label{eq:refMtx}
    \boldsymbol{\Gamma}(\boldsymbol{\vartheta}, t) = \alpha(\boldsymbol{\vartheta}) \;  \mathrm{diag}\left(\left[e^{j\beta_0(t)}, \cdots,  e^{j\beta_{M_u-1}(t)}\right]\right),
\end{equation}
where $\alpha(\boldsymbol{\vartheta}) = 2(2q+1) \cos^{2q}\left(\pi/2 - \arcsin\left(\cos\theta\sin\phi\right)\right)$, modeled as in \cite{9569465}, with $q$ being a parameter that depends on the size of the RIS elements.
Notice that, although $\boldsymbol{\Gamma}(\boldsymbol{\vartheta}, t)$ is a time-varying function, the relation with the incident signal $\mathbf{H}_i(t) * \mathbf{x}(t)$ is multiplicative because $\boldsymbol{\Gamma}(\boldsymbol{\vartheta}, t)$ is the reflection coefficient. It is worth underlining that, for a proper operation of the wireless communication system, the knowledge of the signal $s_d(t)$ is required at the MU side, as the MU must remove $s_d(t)$ from $y_u(t)$ to obtain $\beta(t)$. For full-duplex systems such as the one in Fig. \ref{fig:refScen}, involving only two terminals, the latter condition is easily achieved, but in other implementations, where the receiver of $\beta(t)$ is not the MU, synchronization techniques shall be applied (not covered here).

\subsection{Channel Model}\label{subsec:chModel}

The $(m,n)$-th entry of any of the channel impulse responses in \eqref{eq:dlSignal}-\eqref{eq:ulSignal} can be written following the millimeter wave cluster-based channel model as follows \cite{rappaport2019wireless}:
\begin{align}\label{eq:channel_model}
    [\mathbf{H}(t)]_{m,n} & = \sum_{p=1}^P \xi_p \,\delta\left(t-\tau_p-\Delta \tau_{p,n}-\Delta \tau_{p,m}\right) 
\end{align}
where \textit{(i)} $P$ is the number of paths, \textit{(ii)} $\xi_p\sim \mathcal{CN}\left(0, \sigma_p^2\right)$ is the scattering amplitude of $p$-th path, such that $\sum_{p=1}^P \sigma_p^2 = 1$, \textit{(iii)} $\tau_p$ is the propagation delay of the $p$-th path from the phase center of the Tx to the phase center of the Rx, \textit{(iv)} $\Delta \tau_{p,n}$ and $\Delta \tau_{p,m}$ are the excess propagation delays due to the position of the $n$-th Tx element and the $m$-th Rx element w.r.t. to their phase centers.

As common for communication systems operating through back-scattering and radar systems \cite{manzoni2022motion}, the two-way channel to/from the STMM is characterized by a single dominant path, i.e., $P=1$ for $\mathbf{H}_i(t)$ and $\mathbf{H}_o(t)$. Hence, we have a common propagation delay $\tau=D/c$, where $D$ is the MU-SU distance, and excess delays at the STMM that depend on the incidence and reflection angles $\theta = \theta_o = \theta_i$, $\phi= \phi_o = \phi_i$, which are coincident for full-duplex settings. The inter-element spacing of Tx and Rx arrays (at both MU and SU) is $\lambda_i/2$, while at STMM is $\lambda_i/4$. 
%
%
%

The propagation path losses in power, $\varrho_d$ and $\varrho_u$ in \eqref{eq:dlSignal}-\eqref{eq:ulSignal} are modeled as \cite{9433568}
\begin{equation}\label{eq:pathloss}
    \varrho_d = \frac{2^4 \pi D^2}{\lambda_i^2}, \quad \quad \varrho_u = \frac{2^{12} \pi D^4}{\lambda_i^4}.
\end{equation}
where we assume that \textit{(i)} the inter-element spacing at the STMM is  $\lambda_i/4$, while for the other arrays is $\lambda_i/2$ and \textit{(ii)} the effective area of the MU/SU arrays (and STMM) is equal to the physical one, which is $N(\lambda_i/2)^2$ and $M_u (\lambda_i/4)^2$ \cite{9433568}.

\section{Temporal Modulation}\label{sec:STCDesign}



%
\begin{figure}[t]
    \centering
    \includegraphics[width=\columnwidth]{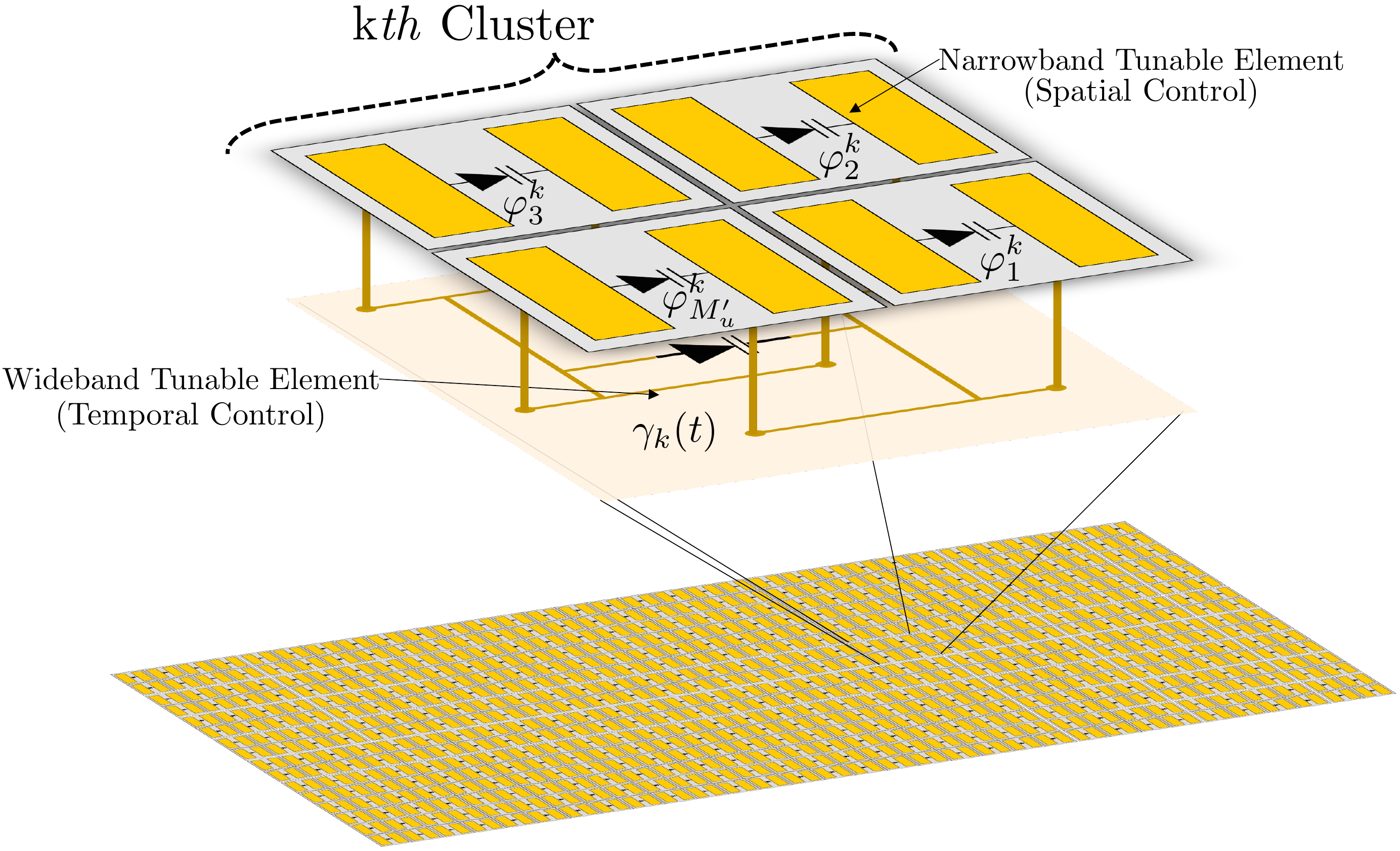}
    \caption{Representation of the proposed time-modulated RIS, with a wideband tunable element (e.g., fast varactor diode) for a cluster of $M_u'$ meta-atoms and narrow band tunable elements (e.g., pin-diodes) for each meta-atom.}
    \label{fig:SPhaseDist}
    \vspace{-0.5cm}
\end{figure}

This section defines the principles of temporal modulation at the STMM and its peculiarities. For the implementation of the STMM, we consider the architecture in Fig. \ref{fig:SPhaseDist}, where the $M_u$ STMM elements are clustered in $K^2$ equal and non-overlapping subsets, each comprising 
\begin{equation}
    M'_{u,x} = \frac{M_{u,x}}{K},\qquad M'_{u,y} = \frac{M_{u,y}}{K}
\end{equation}
elements each along $x$ and $y$. Within the $k$-th cluster, $k=1,...,K^2$, the elements can generate a spatial phase component (different for each element) and a temporal phase component (common to all elements). Considering $K^2=1$ and $K^2=M_u$ leads to the architectures described in our previous work \cite{Mizmizi2023_STMM}, that is here generalized to a hybrid STMM. Reducing $K^2$ means minimizing the cost/complexity of the STMM, while increasing $K^2$ leads to a flexible but expensive architecture (see \cite{rafique2023reconfigurable} for further details). To ease the derivation while maintaining the generality, we consider the MU to use precoding and combining vectors perfectly aligned with the SU. Thus, we have that $\mathbf{f}_d$ and $\mathbf{w}_u$ are given, e.g., estimated by conventional approaches~\cite{channelMiz}.

\subsection{Space-Time Phase Coupling} \label{subsect:SGradientDistortion}

Let us consider the phase signal \eqref{eq:refMtx} applied at the $(q,v)$-th element of the STMM as the superposition of a spatial and a temporal component:
\begin{equation}\label{eq:STPhase}
    \beta_{q,v}(t) = \varphi_{q,v} + \gamma(t),
\end{equation}
for $q=0,..., M_{u,x}-1$, $v=0,..., M_{u,y}-1$, where $\varphi_{q,v}$ is the space component of the phase used to design a complete back-reflection of the impinging signal and $\gamma(t)$ is an information-bearing phase signal applied at each STMM element, and common to all. Here, we assume to apply the same $\gamma(t)$ across the entire STMM to describe the coupling between the spatial and temporal phase gradients. 
\begin{figure*}[h!]
\begin{equation}\label{eq:rxSignalExpanded}
\begin{split}
    y_u(t) & = \rho \; \alpha(\theta,\phi) \,\sum_{q,v}  e^{-j\left(\pi (q \cos \theta \cos \phi + v \cos \theta \sin \phi) -\varphi_{q,v}-\gamma(t-\Delta \tau_{q,v}-\tau)\right)\, h_u(t)} s_d(t-2\tau) + z(t) = \rho\, h_u(t) s_d(t-2\tau) + z(t)  \overset{(a)}{\approx}\\
    & \overset{(a)}{\approx} \rho \; \alpha(\theta,\phi) \sum_{q,v} e^{-j\left(\pi (q \cos \theta \cos \phi + v \cos \theta \sin \phi) -\varphi_{q,v}\right)} e^{j\gamma(t-\tau)} s_d(t-2\tau) + z(t)
\end{split}
\end{equation}
\hrulefill
\end{figure*}
The model of the Rx signal at the MU, reported in \eqref{eq:rxSignalExpanded} can be obtained by substituting \eqref{eq:STPhase} in \eqref{eq:ulSignal}, using the channel model \eqref{eq:channel_model} and assuming that there are no spatially wideband effects at the STMM due to the MU $\rightarrow$ SU signal, namely $s_d(t- 2\Delta T) \approx s_d(t)$, where $\Delta T$ is the maximum excess delay across the STMM elements.  

In \eqref{eq:rxSignalExpanded}, factor $\rho= N^2 \xi_i\xi_o/\sqrt{\varrho_u}$ accounts for geometrical energy losses, while
\begin{equation}\label{eq:propagationDelay_2D}
    \Delta \tau_{q,v} = q \underbrace{\frac{\lambda_i}{4} \frac{\cos \theta \cos \phi}{c}}_{\Delta \tau_x} + v \underbrace{\frac{\lambda_i}{4} \frac{\cos \theta \sin \phi}{c}}_{\Delta \tau_y}
\end{equation}
is the excess propagation delay for the $(q,v)$-th STMM element. 
%
In this setting, unwanted beam squinting from spatially wideband effects due to $s_d(t)$ can be avoided (see \cite{9399122} for details) and the MU $\rightarrow$ SU signal undergoes a \textit{multiplicative} channel $h_u(t)$ that encodes the MU $\leftarrow$ SU phase signal $\gamma(t)$. As demonstrated in \cite{Mizmizi2023_STMM}, it is 
\begin{equation}\label{eq:theorem1}
    \mathbb{E}_\gamma\left[\lvert h_u(t) \rvert^2 \right] \leq M_u^2.
\end{equation}
except for $\gamma(t) = \mathrm{const}$ (straightforward condition) or $\Delta \tau_x = \Delta \tau_y=0$ (perpendicular incidence on the STMM, i.e., $\theta = \pi/2$). This effect is referred here as \textit{space-time phase coupling} and detrimentally affects the Rx signal $y_u(t)$, as shown in Section \ref{subsect:SGradientDistortion}.

Approximation $(a)$ in \eqref{eq:rxSignalExpanded} is subtends that the MU $\leftarrow$ SU phase-only signal does not suffer the relative propagation delay across the STMM, thus $\gamma(t-\Delta T )\approx \gamma(t)$, or equivalently, $\Delta T \ll T_u$
where $T_u$ is the symbol period of the MU $\leftarrow$ SU phase signal $\gamma(t)$. Approximation $(a)$ is typically assumed in the existing literature on STMM, but it is valid in restricted settings (e.g., $\theta = \pi/2$ $\phi = 0$, as in \cite{9133266}). Under $(a)$, the bandwidth of the MU $\leftarrow$ SU Rx signal signal $y_u(t)$ in \eqref{eq:ulSignal} is 
\begin{equation}
    B_{tot} = B_d + B_u
\end{equation}
with $B_u$ being the bandwidth of $e^{j\gamma(t)}$, which is a generic phase modulated signal \cite{anderson2013digital}. Notice that the spectrum of $y_u(t)$ is centered around the carrier frequency $f_i$. This holds for common information-bearing signals $\gamma(t)$ whose average time-derivative over time is zero (thus the net frequency shift is zero). In general, however, the coupling between spatial and temporal phase patterns for $\theta \neq \pi/2$ cannot be neglected for arbitrary MU $\leftarrow$ SU bandwidths, and being peculiar to STMM-based wireless systems, is analyzed herein.

%
%
%
%

\subsection{Effect of Space-Time Phase Coupling} \label{subsect:SGradientDistortion}

\begin{figure}[t!]
\vspace{-.5cm}
    \centering
    \includegraphics[width=0.4\textwidth]{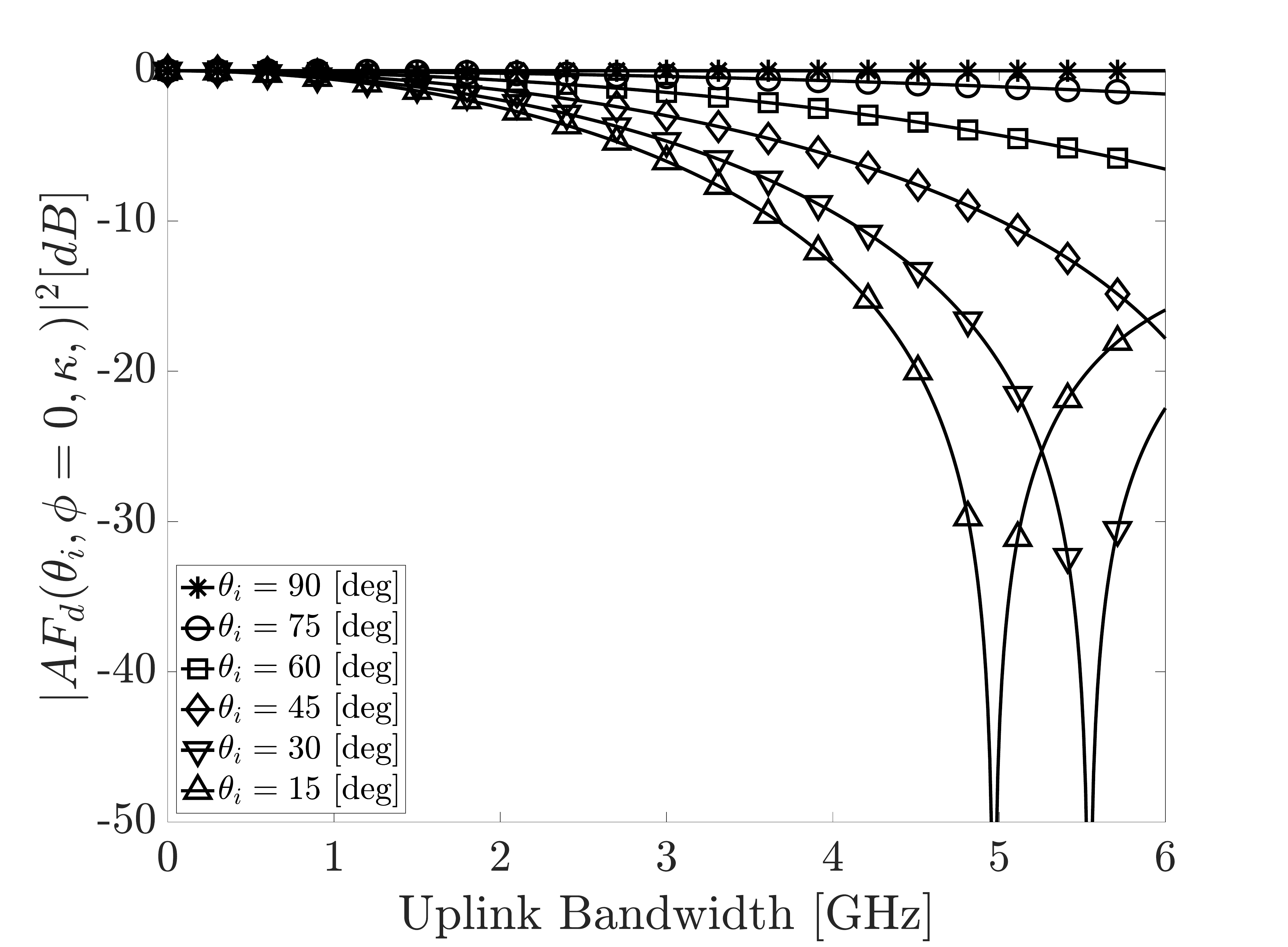}
    \caption{Normalized reflection loss in presence of space-time phase coupling considering different angles of incidence and an STMM size of $100\times100$ elements.}
    \label{fig:NRG}
    \vspace{-.5cm}
\end{figure} 

Inequality \eqref{eq:theorem1} implies that any phase-encoded modulation signal $\gamma(t)$ leads to a decrease in the reflection gain, except for very low MU $\leftarrow$ SU signal bandwidth (approximation $(b)$ in \eqref{eq:rxSignalExpanded}) or $\theta = \pi/2$. The effect of such a space-time coupling phenomenon is due to an unwanted beam squinting, which can be demonstrated by considering the first-order Taylor expansion of $\gamma(t)$ around $t=t_0$:
\begin{equation}\label{eq:GammaLinear}
\begin{split}
    \gamma(t_0-\Delta \tau_{q,v}) \approx \gamma(t_0) -\Dot{\gamma}(t_0) \Delta \tau_{q,v}.
\end{split}
\end{equation}
where $\Dot{\gamma}(t)$ is the time derivative of $\gamma(t)$. The second term is a linear phase shift that causes a drift in the angle of maximum reflection w.r.t. the desired one $(\theta,\phi)$. To quantitatively gain insight into this phenomenon, let us consider a simple linear phase shift $\gamma(t) = 2\pi f_s t$ (e.g., as for one symbol of a frequency shift keying (FSK) modulation), where $f_s = f_i - f_o = \kappa f_i$ is the imposed frequency shift. The normalized array factor of the STMM is
\begin{equation}\label{eq:AF_drift}
    AF(\theta, \phi, \kappa) = \frac{1}{M_u} \sum_{q=0}^{M_{u,x}-1} \sum_{v=0}^{M_{u,y}-1} e^{j 2 \pi f_i \kappa (t-\Delta t_{q,v})}.
\end{equation}
%
%
Figure \ref{fig:NRG} depicts the normalized reflection loss varying the MU $\leftarrow$ SU signal bandwidth $B_u$ and considering different azimuth angles of incidence, and assuming the elevation as $\phi = 0$. As can be observed, when the bandwidth $B_u$ increases for incidence angles $\theta_i \neq \pi/2$, the space-time coupling will cause a loss in the reflection gain and subsequently in the performances of the system.

The loss observed in Fig. \ref{fig:NRG} is due to the tilting of the angle of maximum reflection, which can be evaluated as
\begin{equation}\label{eq:spatialDrift}
    \overline{\theta} = \arccos \left[(1 + \kappa) \cos \theta \right]
\end{equation}
where $\overline{\theta} \neq \theta$ except for $\kappa=0$. Notice that for $|(1 + \kappa) \cos \theta | > 1$ in \eqref{eq:spatialDrift} we have evanescent waves. 
It is clear that to implement a wideband wireless communication system using STMM, the spatial phase must be decoupled from the temporal one, as described in subsection \ref{subsect:Decoupling}.

\subsection{Space-Time Phase Decoupling}\label{subsect:Decoupling}
Decoupling the spatial phase from the temporal one can be achieved differently depending on the specific STMM architecture. In the general case in which $K^2 < M_u$ (and often $K^2 \ll M_u$), the decoupling is made for the $k$-th cluster of elements as follows:
\begin{equation}\label{eq:CDecoupling}
    \beta_{q,v}(t) =  \varphi_{q,v} + \gamma(t + \Delta \tau_{k}) - \Dot{\gamma}(t + \Delta \tau_{k})\left(\Delta \tau_{q,v} + \Delta \tau_{k}\right)
\end{equation}
where $\Delta \tau_{k}$ is the excess propagation delay for the $k$-th cluster center w.r.t. the STMM phase center. If $K^2=M_u$, the decoupling simplifies to $\beta_{q,v}(t) = \varphi_{q,v} + \gamma(t + \Delta \tau_{q,v})$.  Expression \eqref{eq:CDecoupling} is derived from \eqref{eq:GammaLinear} applied to the $k$-th cluster, and allows compensating for the first time-derivative of the information signal $\gamma(t)$, i.e., the instantaneous frequency shifts.

\section{Numerical Results}\label{sect:FDComm}


%
%


We show numerical results illustrating the effects of space-time phase coupling on the MU $\leftarrow$ SU spectral efficiency, and the benefits of the proposed decoupling. The goal is to provide insights on upper achievable performance, operating conditions, and possible limitations. In all the following, the frequency of operation is $f_0=30$ GHz, link length is $D = 100$ m, transmitted power by the MU is $20$ dBm, the noise power spectral density is $N_0 = -173$ dBm/Hz, and the STMM size is $100\times100$ elements. We consider a continuous phase-frequency shift keying (CP-FSK) modulation with modulation index $h=1$. The $99\%$ of the signal energy is within $B_u = 2/T_u$ \cite{kuo2004bandwidth}, and a fixed total system bandwidth of $B_{tot} = 5$ GHz, split between $B_d$ and $B_u$. The trends and considerations are the same for a generic phase/frequency modulation, but the absolute values must be derived case by case.

We evaluate the total unconstrained spectral efficiency of the MU $\rightleftarrows$ SU link, that can be expressed as follows:
\begin{equation}\label{eq:Ucapacity}
\begin{split}
        \eta = \mu_u \log_2(1+\mathrm{SNR}_u) + (1-\mu_u) \log_2(1+\mathrm{SNR}_d)
\end{split}
\end{equation}
where $\mu_u = B_u/B_{tot}$, while $\mathrm{SNR}_u$ and $\mathrm{SNR}_d$ are the MU $\leftarrow$ SU and MU $\rightarrow$ SU signal-to-noise ratio (SNR) at the decision variable, respectively. 
The $\mathrm{SNR}_d$ is generally expressed as
\begin{equation}\label{eq:snr_d}
    \mathrm{SNR}_d \leq \frac{\sigma_{s,d}^2 N M_d}{\varrho_d \sigma_{z}^2}
\end{equation}
while $\mathrm{SNR}_u$ is upper-bounded by
\begin{equation}\label{eq:snr_u}
    \mathrm{SNR}_u \leq \frac{\sigma_{s,d}^2 N^2 M_u^2}{\varrho_u \sigma^2_{z}}  |AF(\theta, \phi, \kappa)|^2 T_u B_{tot}
\end{equation}
where the equality holds when perfect channel state information is available and optimal precoding/combing is employed, which both provide the maximum MIMO gain, i.e., $N M_d$, and $|AF(\theta, \phi, \kappa)|^2 \leq 1$ is the reflection loss due to the space-time phase coupling. The expression \eqref{eq:snr_u} is obtained considering a matched filtering approach at the MU side over the MU $\leftarrow$ SU symbol duration $T_u$, aimed at estimating the phase signal $\gamma(t)$. The matched filter gain is $T_u B_{tot}$. For the system depicted in Fig. \ref{fig:refScen}, we can assume the MU knows the MU $\rightarrow$ SU signal $s_d(t)$ and has correctly performed the timing recovery to estimate the macroscopic delay $\tau$. The analytical expression of the true $\mathrm{SNR}_u$ is derived in our work \cite{Mizmizi2023_STMM}.


\begin{figure}[t!]
\vspace{-.5cm}
    \centering
    \includegraphics[width=0.45\textwidth]{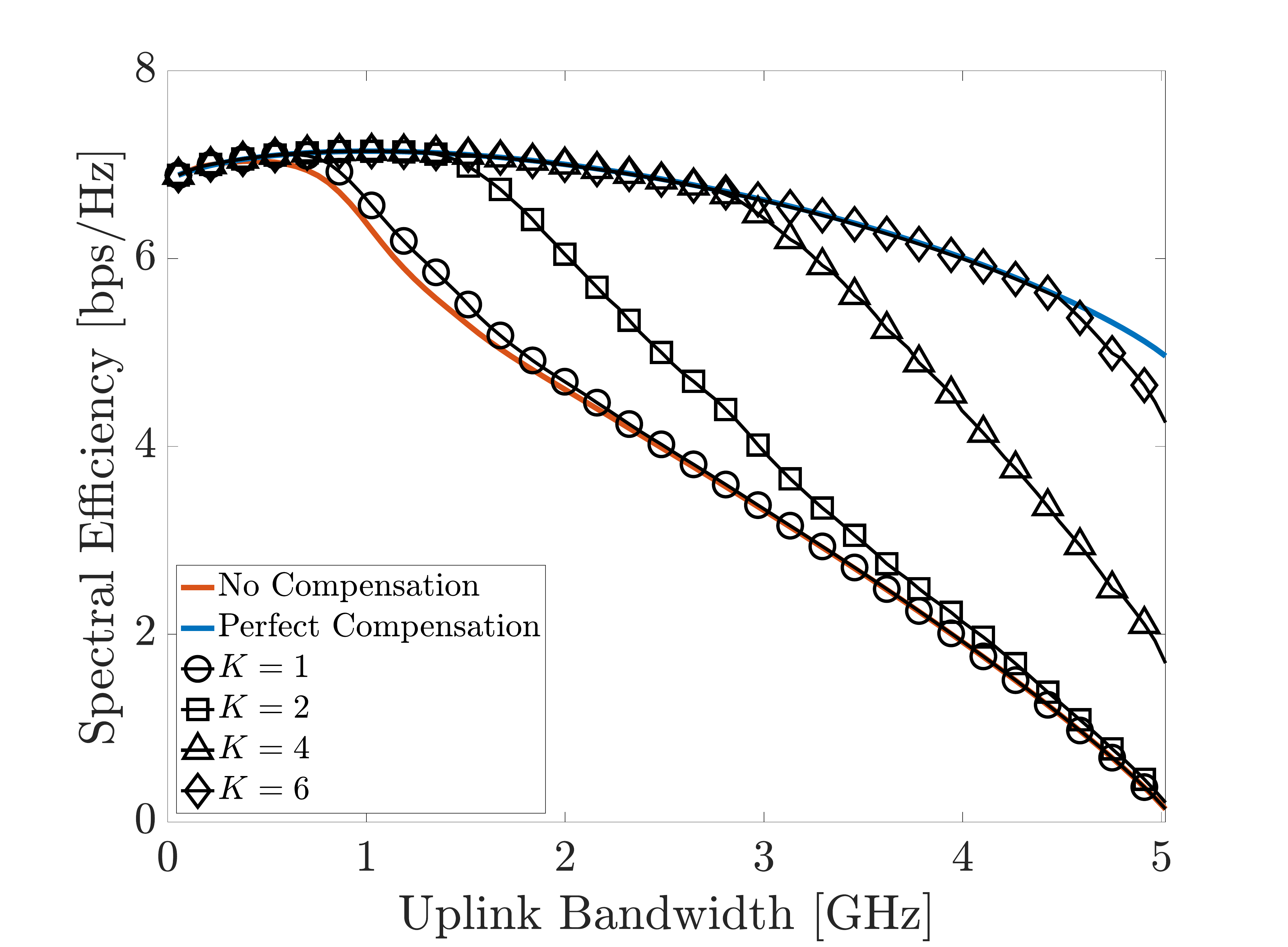}
    \caption{Total Achievable spectral efficiency varying the uplink bandwidth $B_u$, for an azimuth incidence angle of $\theta_i = 30$ [deg], and considering a different number of clusters $K$.}
    \label{fig:SEvsB}
    \vspace{-.4cm}
\end{figure} 
\begin{figure}[t!]
    \centering
    \includegraphics[width=0.45\textwidth]{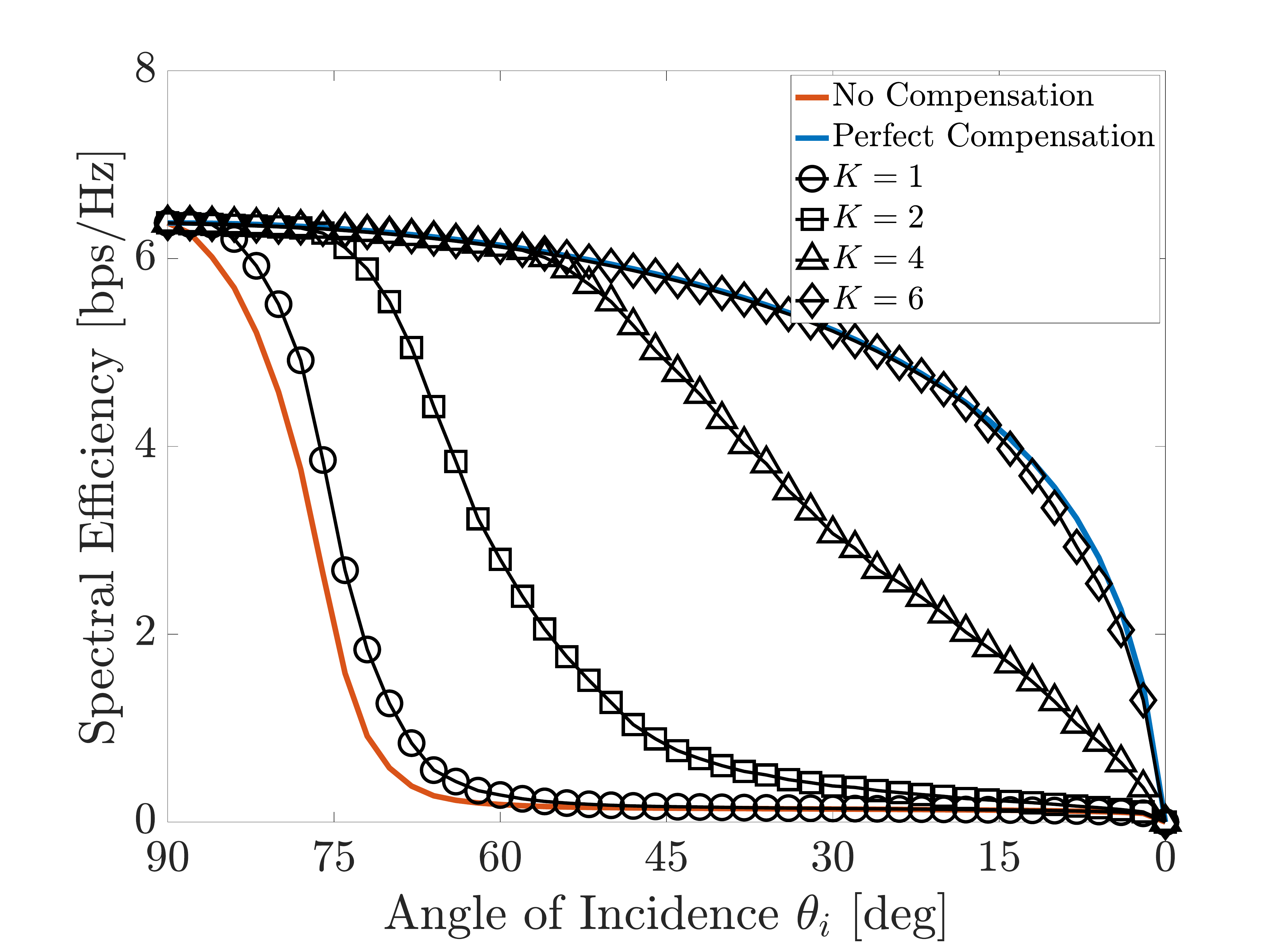}
    \caption{Total Achievable spectral efficiency varying the azimuth incidence angle $\theta_i$, for an uplink bandwidth of $B_u = 4$ GHz, and considering a different number of clusters $K$.}
    \label{fig:SEvsAz}
    \vspace{-.5cm}
\end{figure}

Figure \ref{fig:SEvsB} depicts the total unconstrained spectral efficiency of the system varying the uplink bandwidth $B_u$. The blue curve shows the performance achieved when perfectly compensating the coupling, i.e., $|AF(\theta, \phi, \kappa)|^2 = 1$, $\forall \theta, \phi$ and $\kappa$. This result can be achieved if $K^2 = M_u$ and by applying \eqref{eq:CDecoupling}. It is interesting to observe that there is an optimal uplink bandwidth $B_{u, opt} \approx 1$ GHz, which depends on the tradeoff between the pathloss $\varrho_u$, the processing gain $T_u B_{tot}$, and the size of the STMM $M_u$. The red curve depicts the system's performance when the space-time coupling is not compensated, which drops dramatically as the uplink bandwidth $B_u$ increases. The black curves show the performance when compensating the space-time coupling with a limited number of clusters $K^2$. The main takeaway is that when the uplink bandwidth $B_u$ increases more clusters $K^2$ are necessary to maintain optimal performances.

The impact of the azimuth angle of incidence $\theta_i$ is depicted in Fig. \ref{fig:SEvsAz} given an uplink bandwidth of $B_u = 4$ GHz. Without space-time coupling compensation (red line), as the angle of incidence becomes grazing we observe a rapid decay in achievable spectral efficiency. The blue line, which refers to the perfectly compensated coupling, is influenced only by the loss due to the reduced reflectivity of the RIS elements, i.e., the term $\alpha(\boldsymbol{\vartheta})$. Again, we observe that as the number of clusters $K$ increases, the space-time decoupling is more robust, which highlights the validity of the method proposed.

\section{Conclusion} \label{sect:conclusion}
The paper analyzes the space-time phase coupling arising in STMM-based wireless communication systems, where a MU equipping a full-duplex transceiver is transmitting an informative signal toward a SU, that back-reflects its information by modulating the impinging signal with the STMM. The analytical findings show that the space-time coupling depends on the direction of incidence, the bandwidth of the SU's informative signal, and the size and type of the STMM. A method for space-time phase decoupling is then proposed and validated through simulations.

\bibliographystyle{IEEEtran}
\bibliography{Bibliography}

\end{document}